\documentclass[aps, twocolumn,showkeys,showpacs,amsmath,amssymb,superscriptaddress]{revtex4-1}
\usepackage{graphicx}
\usepackage{rotating}
\usepackage{latexsym}

\newcommand{\beq}{\begin{equation}}
\newcommand{\eeq}{\end{equation}}
\newcommand{\bqa}{\begin{eqnarray}}
\newcommand{\eqa}{\end{eqnarray}}

\begin{document}
\title{A fully relativistic lattice Boltzmann algorithm}

\author{P. Romatschke} \email{romatschke@fias.uni-frankfurt.de}
\affiliation{ Frankfurt Institute for Advanced Studies, D-60438
  Frankfurt (Germany) }

\author{M. Mendoza} \email{mmendoza@ethz.ch} \affiliation{ ETH
  Z\"urich, Computational Physics for Engineering Materials, Institute
  for Building Materials, Schafmattstrasse 6, HIF, CH-8093 Z\"urich
  (Switzerland)}

\author{S. Succi} \email{succi@iac.cnr.it} \affiliation{Istituto per
  le Applicazioni del Calcolo C.N.R., Via dei Taurini, 19 00185, Rome
  (Italy),\\and Freiburg Institute for Advanced Studies,
  Albertstrasse, 19, D-79104, Freiburg, Germany}

\date{\today}
\begin{abstract} 
Starting from the Maxwell-J\"uttner equilibrium distribution,
we develop a relativistic lattice Boltzmann (LB)
algorithm capable of handling ultrarelativistic systems with
flat, but expanding, spacetimes.  
The algorithm is validated through simulations of quark-gluon
plasma, yielding excellent agreement with hydrodynamic simulations.
The present scheme opens the possibility of transferring the recognized 
computational advantages of lattice kinetic theory to the context of
both weakly and ultra-relativistic systems.
\end{abstract}



\maketitle

\section{Motivation}

The great success of the Relativistic Heavy-Ion Collider (RHIC)
experimental program
\cite{Adcox:2004mh,Back:2004je,Arsene:2004fa,Adams:2005dq} has
provided the motivation to come up with realistic and quantitative
simulations of heavy-ion collisions.

Since the bulk of particles produced in relativistic heavy-ion
collisions is described by fluid dynamics \cite{Kolb:2003dz}, the
center-piece of any complete simulation attempt will involve a viscous
fluid dynamics algorithm.  The majority of presently available fluid
dynamics codes is able to handle smooth initial conditions in 2+1
dimensions in the presence of shear
viscosity\cite{Romatschke:2007mq,Dusling:2007gi,Song:2007ux,Chaudhuri:2008je,Niemi:2011ix}.
However, it has by now been understood that the presence of
event-by-event fluctuations in the initial state can lead to
significantly different quantitative results with respect to smooth
initial conditions \cite{Nagle:2010zk}, and may in some cases even
explain qualitatively new phenomena. To be more specific, the presence
of event-by-event fluctuations is the source of the non-vanishing
elliptic flow found in RHIC experiments at central collisions, the
source of hydrodynamic flow-fluctuations, 
and may (through the presence of so-called triangular flow $v_3$) 
naturally explain the
presence of the 'ridge phenomenon' found in experiments
\cite{Teaney:2010vd,Luzum:2010fb,Luzum:2010sp,Sorensen:2011hm}.  Thus,
it is probably fair to say that without including the effect of
event-by-event fluctuations, a description of the medium created in
heavy-ion collisions cannot be regarded as realistic. This provides
the motivation to develop a fully relativistic and computationally efficient 
viscous fluid dynamics algorithm that can handle initial state fluctuations.
Also, such an algorithm can be used to validate the only available 3+1 dimensional
relativistic viscous hydrodynamics code by the McGill group \cite{Schenke:2010rr}.

Further motivation is provided by other systems where relativistic viscous
fluid flows are of interest, such as astrophysical systems and
condensed matter systems such as graphene \cite{Graphene}. One
particular question that arises in all this different systems is when
relativistic fluid flow becomes turbulent, which involves a
determination of the critical Reynolds number and the turbulent
spectrum \cite{Romatschke:2007eb,Fouxon:2009rd}.

\section{Lattice kinetic approach to hydrodynamics}

Fluid turbulence, both classical and relativistic, sets one of the most
compelling challenges in modern computational physics.  
This motivates a relentless search for new and ever more efficient methods for
solving the hydrodynamic equations of motion in the high-Reynolds turbulent
regimes.  
In the last two decades, a new computational paradigm has emerged, which is
based on the idea of solving hydrodynamic problems from the standpoint
of Boltzmann kinetic theory. Apparently, this is rather counterintuitive,
because the Boltzmann equation lives in a double-dimensional (phase)
space, consisting of three dimensions in ordinary space, plus three
additional dimensions in momentum (velocity space). In addition, the
Boltzmann equation contains the details of microscopic interactions
through a very complicated integral collision operator in velocity
space, which is computationally very demanding. As a result,
the Boltzmann equation has never been considered a practical tool for
computational fluid dynamics, apart from the special case of rarefied
gas dynamics, for which ordinary fluid dynamics is known to be
inadequate. In the last two decades, however, minimal realizations of
the Boltzmann equation have been developed, which relinquish the
aforementioned problems, and gave rise to a computational method of
remarkable elegance and outstanding computational efficiency.  Since
these minimal forms of Boltzmann kinetic equations are formulated in a
discrete velocity and space-time lattice, they have come to be known
as Lattice Boltzmann Equation(s) (LBE) \cite{LBE1, LBE2, LBE3}.  
To date, LB methods have met with amazing success across virtually all
sectors of {\it non-relativistic} fluid dynamics, from flows to porous
media, to turbulent flows in complex geometries, multiphase,
colloidal, hemodynamic flows, and magnetohydrodynamics \cite{LBEALL1,
  LBEALL2, LBEALL3, LBEALL4, LBEALL5, LBEALL6, LBEALL7, LBEALL8}.
However, relativistic formulations have come into existence only very 
recently \cite{RELB1, RELB2}.  
Indeed, the lattice formulation of the relativistic Boltzmann equation presents a series
of theoretical and methodological challenges which have no counterpart in
the non-relativistic realm.  To date, some of these challenges have
been bypassed by formulating a relativistic LBE (RLBE) {\it top-down},
i.e.  by recovering the equations of relativistic hydrodynamics
through a moment-matching procedure in the lattice.  
This gives rise to a very efficient computational scheme for mildly relativistic
systems, but does not guarantee the so-called "realizability" of RLBE, i.e.  the 
fact that RLBE should be derived by an underlying microscopic model, or, at
least, by a continuum version of a relativistic kinetic equation.
Even leaving aside computational considerations, this is an important task
in the process of placing the RLBE onto a solid conceptual framework.
This is precisely the task accomplished in the present paper.  

\section{Relativistic kinetic theory}

Before discussing the details of LB methods,
let us briefly review the theoretical background of kinetic theory in a relativistic context.
The starting point is the Boltzmann equation for the single particle
distribution $f = f(x^{\mu},p^{\alpha})$, with the relativistic
analogue of the Bhatnagar-Gross-Krook collision term\cite{CerREL}:
\beq \left[p^\mu \nabla_\mu - \Gamma^\lambda_{\mu\nu} p^\mu p^\nu
  \partial_\lambda^{(p)} \right] f = - \frac{p^\mu
  u_\mu}{\tau_R}\left(f-f^{\rm eq}_J\right)\,,
\label{BE}
\eeq 
Here $x^{\mu}$ and $p^{\alpha}$ are the position and momentum
4-vectors, respectively, $\tau_R$ is the single relaxation time, and
$f^{\rm eq}_J$ denotes the equilibrium distribution function. Here and in the following,
we work in units where the speed of light $c$, the Boltzmann constant $k_B$ and
Planck's constant $\hbar$ have been set to unity, $c=k_B=\hbar=1$.
In the ultrarelativistic case, this may be taken in the form of the J\"uttner
distribution function
\beq f^{\rm eq}_J = Z^{-1} e^{-p_\mu u^\mu/T}\,,
\label{Juttner}
\eeq 
where $Z^{-1}$ parametrizes the number of degrees of freedom,
$u^\mu$ is the macroscopic (fluid) 4-velocity, $T$ is the
local temperature, $\nabla_\mu$ denotes the (geometric) covariant
derivative, and $\Gamma_{\mu \nu}^\lambda$ are the Christoffel symbols
that are given by derivatives of the underlying metric tensor
$g_{\mu\nu}$.

The connection to fluid dynamics is realized by introducing the
energy-momentum (energy-stress) tensor $T^{\mu\nu}$,
\bqa 
T^{\alpha \beta}(t,x)  &\equiv& \int d\chi p^\alpha p^\beta f(t,x,p)
\\ \nonumber &\equiv& \int \frac{d^4p}{(2\pi)^3} \delta(p^\mu p_\mu -m^2) 2 H(p^0)
p^\alpha p^\beta f(t,x,p) \,, 
\eqa
where $m$ is the particle mass and $H$ the Heavyside step function. 
Note that we have introduced the notation $f(t,x,p)  \equiv f(x^\mu,p^\alpha)$.
The equations of motion obeyed by the tensor $T^{\mu\nu}$ emerge, after a little
algebra, upon integrating Eq.~(\ref{BE}) with respect to four-momentum degrees
of freedom, $\int d\chi p^\nu$, 
\beq
\label{eom1}
\nabla_\mu T^{\mu\nu}=-u_\mu \int d\chi p^\mu p^\nu \frac{\left(f-f_{\rm eq}\right)}{\tau_R}=
- \frac{u_\mu}{\tau_R}\left(T^{\mu\nu}-T^{\mu\nu}_{\rm eq}\right)\,.
\eeq

The equilibrium energy-momentum tensor $T^{\mu\nu}_{\rm eq}$ is readily
computed using the J\"uttner distribution, and reads as follows:
$$
T^{\mu \nu}_{\rm eq}=(\epsilon+P) u^{\mu} u^\nu - P g^{\mu\nu}\,,
$$
where the energy density $\epsilon$ and pressure $P$ are functions of
the temperature and the number of degrees of freedom $Z$ (here arbitrarily set
to one). The full energy momentum tensor may then be written as
$T^{\mu\nu}=T^{\mu\nu}_{\rm eq}+\Pi^{\mu\nu}$, where the second
term collects non-equilibrium contributions.
  
Still needed is the choice of the rest-frame of the heat bath with respect to which the
fluid velocity $u^\mu$ is defined. In the ultrarelativistic limit, the
canonical choice is the so-called Landau-Lifshitz condition, whereby
$u_\mu T^{\mu\nu}\equiv \epsilon u^\nu$, so that Eq.~(\ref{eom1})
reads simply as:
$$
\nabla_\mu T^{\mu\nu}=0\,,
$$
expressing the (covariant) conservation of energy and momentum.  
Sufficiently close to equilibrium, where gradients are small, the form of $\Pi^{\mu\nu}$
can be calculated by integrating Eq.~(\ref{BE}) with
respect to $\int d\chi p^\mu p^\alpha$
(c.f.~\cite{Baier:2006um,Baier:2007ix}). 
In the ultrarelativistic case, when particle masses can be neglected ($m=0$) one finds:
$$
\Pi^{\alpha\beta}=  \tau_R \frac{\epsilon+P}{6}\nabla^{<\alpha} u^{\beta>}\,,
$$
where
$$
A^{<\alpha}B^{\beta>} \equiv \left( \Delta^{\alpha \mu}\Delta^{\beta\nu}
+\Delta^{\alpha \nu}\Delta^{\beta\mu}-\frac{2}{3}\Delta^{\alpha
 \beta} \Delta^{\mu \nu}\right)A_\mu B_\nu \,,
$$
and $\Delta^{\mu\nu}=g^{\mu\nu}-u^\mu u^\nu$. Performing a
non-relativistic limit of $\nabla_\mu T^{\mu\nu}$ one recovers the
Navier-Stokes equations, with the following dynamic viscosity coefficient
$$
\eta = \tau_R \frac{\epsilon+P}{6}\,.
$$

Therefore, Eq.~(\ref{BE}) reproduces the equations of fluid dynamics
in the continuum, on condition that $\tau_R=6 \frac{\eta}{s} T^{-1}$, where $s$ denotes 
the entropy density and gradients must be small
enough that a fluid dynamics description makes sense at all.
A few remarks are in order: by construction, kinetic theory achieves a
remarkable disentangling between non-linearity and non-locality, which
proves beneficial for both theoretical and computational purposes.  
Indeed, in the hydrodynamic formulation, any generic quantity, including the flow
velocity, is transported along space-time changing trajectories,
defined by the flow velocity itself, thereby giving rise to terms which
are non-local and non-linear at a time. 
In a turbulent flow, such trajectories become typically fairly complicated, thus opening
potential exposures to numerical inaccuracies and instabilities.  
In kinetic theory, on the contrary, information is always transported
along constant characteristics, $dx=vdt$, since the velocity $v$
(vector notation relaxed for simplicity) does not depend on space-time
coordinates. Thus, no matter how wild the space-time dependence of the 
fluid flow, the streaming operator is linear in the Boltzmann
distribution function, and the information always travels along straight lines.  
The price to pay for this major advantage is the need of three
extra-dimensions in velocity space. However, velocity space lends itself to very economic
discretizations, typically $O(3^d)$ discrete velocities in $d$ spatial dimensions space,
which make the tradeoff between non-linearity and over-dimensionality an excellent bargain,
and is one of the key assets of the LB formulation.

One may then wonder where the non-linearity has disappeared in the
kinetic formulation.  It turns out that it is 
concealed in the local equilibrium distribution, which is a non-linear function of
the local hydrodynamic variables, see expression (\ref{Juttner}). 
Furthermore, note that since collisions are zero-ranged in the Boltzmann
picture, the corresponding collision term is completely local, as
anticipated.  This lies at the root of the excellent amenability of LB
to parallel computing, another major practical asset of the technique
altogether \cite{LBEALL2}.

Having clarified the main philosophy of the kinetic pathway to
fluid dynamics, one must come down to specific details. 
The main question is: how sparse can the sampling in
momentum space be made? 

The target criterion is that $\nabla_\mu T^{\mu \nu}=0$ must emerge 
as a continuum limit of the {\it lattice analogue} of Eq.~(\ref{BE}). 
Like in the non-relativistic framework, this sets a specific demand
on the symmetry of the lattice tensors, as detailed in a sequel to this work.
In particular, second order tensorial identities of the form
\beq\label{tmunucons}
u_\mu \int d\chi p^\nu p^\mu f(t,x,p) =\epsilon\,u^\nu \,.  
\eeq
have to be reproduced exactly in the lattice formulation.
\section{Detour: the non-relativistic Lattice Boltzmann formulation}
At this point, it is instructive to review the setup of the LB scheme
in the non-relativistic context.  There, the equilibrium
distribution function for an ideal gas is the Maxwell distribution,
$$
f^{\rm eq}_M (t,x,v)= \frac{\rho(t,x)}{\sqrt{2 \pi c_{s}^2}}
e^{-\frac{({\bf v}-{\bf u})^2}{2 c_s^2}} \,,
$$
where $c_s =\sqrt{T/m}$ is the sound speed, the velocities ${\bf v}$
(alternatively denoted as $v^i$ with $i$ running on spatial coordinates) take the role of the momenta
$p$, and ${\bf u}$ is the macroscopic velocity. 
By introducing a scaled temperature $\theta=\frac{T}{T_0}$, $T_0$ being a reference
temperature, and rescaling the velocities with the
speed of sound, the velocity moments (the
conditions to recover fluid dynamics corresponding to
Eq.~(\ref{tmunucons})), take the form \cite{Shan:1997}
\beq
\label{velmoms}
\int d^3v f(t,x,v) =\rho\,, \qquad \int d^3v {\bf v} f(t,x,v) = \rho
{\bf u},
\eeq
$$ 
\int d^3v {\bf v}^2 f(t,x,v) = 2 \rho \epsilon_{\rm
  int}+\rho {\bf u}^2\,,
$$
where $\rho$ is the mass density and $\epsilon_{\rm int}$ the internal
energy density. Note that the local Maxwellian corresponds to
the generating functional of the Hermite polynomials $H_n(v)$, \beq
\label{Hermitgen}
e^{-(v-u)^2/2}=e^{-v^2/2}\sum_{n}H_{n}(v)\frac{u^n}{n!}\,, \eeq in
one-dimension. 
This is readily generalized to three dimensions, thanks to the factorizability
of the pre-factor $e^{-{\bf v}^2/2}$ into the respective components. 
The actual Maxwellian distribution can then be approximated as
$$
e^{-(v-u)^2/2\theta} \sim e^{-v^2/2\theta}\sum_{n=0}^N
\frac{a_n(u/\theta^{1/2})}{n!}H_{n}(v/\theta^{1/2})\,,
$$
which is valid in the sense of mean convergence, if
$$
\int d{\bf v} e^{v^2/2} e^{-|{\bf v-u}|^2/2\theta} \,,
$$
exists, or equivalently $\theta<2$ (c.f. \cite{Grad,Shan:1997}).  
The same representation can be applied to the
full distribution $f(t,x,v)$, which reads (after rescaling $v \rightarrow v\, \theta^{1/2}$, $u \rightarrow u\, \theta^{1/2}$):
\beq
\label{nonrelf}
f(t,x,v)=e^{-v^2/2}\lim_{N\rightarrow \infty}\sum_{n=0}^N\frac{a_n^{i_1\ldots
 i_n}(t,x)}{n!}  {\cal H}_n^{i_1\ldots i_n}({\bf v})\,,
\eeq
where ${\cal H}_n^{i_1\ldots i_n}$ are tensor Hermite polynomials
(c.f. \cite{Grad}).  In practice, $f$ will be approximated by truncating the above
sum at finite (small) $N$.

\begin{figure}[t]
\center
\includegraphics[width=0.9\linewidth]{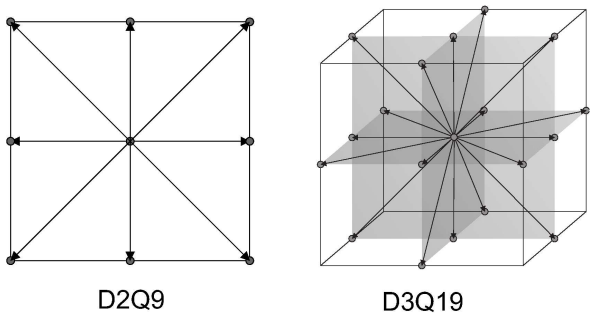}
\hfill
\caption{Typical lattice configurations D2Q9 (9 velocities
in 2 dimensions) and D3Q19 (19 velocities in 3 dimensions)
for the lattice Boltzmann model. 
The arrows denote the discrete unit vectors set.
}
\label{d2q9}
\end{figure}

Integrals of the form $\int e^{-v^2/2} P({\bf v}) d {\bf v}$,
where $P$ is a polynomial of degree $2 N$ or less, can then be
calculated \emph{exactly} as a sum over the roots of $H_N(v)$ (``Gauss-Hermite quadrature'').  
Therefore, the roots $v=v_m$, $m=1,\ldots N$ represent the ideal 
choice for the discretization/sampling of velocity space.  
To guarantee that the Boltzmann equation reproduces the non-relativistic fluid dynamics
equations, the velocity moments must be represented exactly, which
implies the necessary condition $N \ge 2$. 
For many applications, specifically those not dealing with strong thermal
and compressible phenomena, this is also sufficient to a second order
numerical accuracy.
The end result is a discrete (lattice) Boltzmann equation of the form
\begin{equation}
  \label{LBE}
  f_i(t + \Delta t, {\bf x}+{\bf c}_i \Delta t)-f_i(t, {\bf x})=-\Omega(f_i-f_i^{eq}) \,,
\end{equation}
where $\Omega = \frac{\Delta t}{\tau_R}$ with $\tau_R$ the single
relaxation time. The discrete velocities ${\bf c}_i$ run
over a lattice with sufficient symmetry to guarantee mass, momentum
and momentum-flux conservation (\ref{velmoms}) , so as to recover the exact form of the
Navier-Stokes equations.  
Typical lattices fulfilling the
above constraints are the D2Q9 (nine velocities in two dimensions) and
D3Q19 (nineteen speeds in three dimensions), see Fig.~\ref{d2q9}.

  

Note that, owing to the Cartesian formulation, these lattice
configurations are space-filling. This is crucial to ensure the so
called {\it light-cone} condition $d {\bf x}_i = {\bf c}_i dt$,
i.e. the discrete populations hop from site to site in fully
synchronous mode (grid-bound dynamics), a feature which turns out to
be critical to the computational efficiency of the LB scheme.  Upon
performing a standard Chapman-Enskog asymptotic expansion, the lattice
Boltzmann scheme (\cite{LBE1, LBE2, LBE3}) is shown to recover the hydrodynamic
equations of a quasi-incompressible fluid with kinematic viscosity (in
lattice units $\Delta t = \Delta x =1$) $\nu=c_s^2(\tau_R-1/2)$, $c_s$
being the lattice sound speed, typically $1/\sqrt 3$.  To be noted,
the factor $-1/2$ at the right hand side, which stems from a second
order Taylor expansion of the discrete streaming LBE operator.  This
term, known as 'propagation viscosity', contributes a negative
viscosity and permits to achieve very small viscosities $\nu \ll 1$
with unit time-steps, by simply choosing $\tau_R = 1/2+\nu/c_s^2$.
This property is crucial to access low-viscous, turbulent regimes
while still preserving an efficient time-marching procedure.

\section{A fully relativistic LB algorithm}

The lattice formulation of relativistic kinetic theory poses a few
genuinely new challenges, primarily the fact that the energy $E$ is
no longer a simple quadratic function of the velocity (momentum)
$E = \sqrt{m^2 + {\bf p}^2}$.
This basic feature reflects in the non-separability
of the J\"uttner distribution along the three components of 
the momentum ${\bf p}$, and forbids a simple transcription
of the Hermite procedure described above for the case of non-relativistic fluids. 

This is the reason why the only existing relativistic LBE version available 
to date is based on a {\it top-down} procedure, i.e. design lattice equilibria with free
Lagrangian parameters, which are then adjusted in such a way as to
match the five basic conservation laws, number density and
energy-momentum 4-vector.  Full details can be found in the original
references \cite{RELB1,RELB2}.  The scheme was validated for two
different relativistic applications, 1d quark-gluon plasmas, 3d
supernova explosions, and graphene \cite{turbPRL}, showing excellent
performance on all of them.  However, inherent to the moment-matching
procedure, is the question of realizability, i.e. the existence of an
underlying microscopic model or at least an equivalent analogue
in continuum kinetic theory.  
Moreover, the top-down procedure only
works in the case of Cartesian coordinates.  
In view of general relativistic applications, involving generic coordinate 
systems, it is highly desirable to explore the viability of the relativistic LB
procedure beyond the Cartesian realm (incidentally, this would prove
useful also for non-relativistic applications in general coordinates).

For the ultra-relativistic case, where particle masses can be neglected,
the equilibrium distribution reads as 
\beq
\label{expp}
e^{-p\cdot u/T}=e^{-|{\bf p}| u^0/T + {\bf p}\cdot {\bf u}/T}\,, \eeq
which does not allow an ansatz such as Eq.~(\ref{Hermitgen}) because
the pre-factor corresponds to an unfactorizable square root dependence.
This alone prevents a straightforward use of Cartesian coordinates.

Rather, the equilibrium distribution function suggests the following
expansion: \beq
\label{fiexp}
e^{-p\cdot u/T}=e^{-|{\bf p}| u^0/(T_0 \theta)} 
\sum_n\left(\frac{\bf
 p}{|{\bf p}|}\right)^n \left(\frac{|{\bf p}| u^0}{T_0
 \theta}\right)^n \frac{({\bf u})^n}{(u^0)^n n!}\,, \eeq which
involves unit vectors ${\bf v}=\frac{{\bf p}}{|{\bf p}|}$ rather than
momenta ${\bf p}$ and powers of $\frac{|{\bf p}| u^0}{T_0\theta}$, that
go together with the exponent and where again the scaled temperature
$\theta=\frac{T}{T_0}$ was introduced. 

This leads to the following ansatz for the relativistic distribution function $f$:

\beq
\label{ansatz}
f(t,x,p)=e^{-p^0/T_0} \sum_n P^{(n)}_{i_1\ldots i_n}({\bf v})
a^{(n)}_{i_1\dots i_n}\left(t, x, p^0/T_0\right) \,, \eeq 
where $p^0=|{\bf p}|$ and vector
polynomials $P^{(n)}_{i_1\ldots i_n}({\bf v})$ which are orthogonal
with respect to the angular integral $\int \frac{d\Omega}{4\pi}$. Their properties are listed in appendix \ref{sec:B}.

The ansatz (\ref{ansatz}) should constitute a valid approximation to
$f(t,x,p)$ in the sense of mean convergence, i.e. provided that the
integral 
$$
\int_0^\infty dp^0 e^{p^0/T_0} f^2(t,x,p) \,,
$$ 
exists, which together with Eq.(\ref{expp}) implies $\theta<2
\left(u^0-|{\bf u}|\right)$. For highly relativistic fluid flows such
as those with Lorentz factor of $ \gamma \sim 10$, $\theta=0.1$ would ensure
convergence. However, in practical applications, such as the dynamics
of heavy-ion collisions, such high fluid flow velocities only occur
for the small temperature region. 
Hence, setting $T_0$ to the maximum temperature encountered in the problem 
was found to give acceptable results.

Denoting the scaled momentum as $\bar{p}=p^0/T_0$, it proves 
convenient to further expand the coefficients $a^{(n)}$ in terms of
generalized Laguerre polynomials $L_k^{(\alpha)}$, so that the
complete ansatz for $f$ is given by \beq f(t, x, p)=e^{-\bar{p}}
\sum_{k=0}^{N_p-1}\sum_{n=0}^{N_v-1} P^{(n)}_{i_1\ldots i_n}({\bf v})
L_{k}^{(\alpha)}\left(\bar{p}\right) a^{(nk)}_{i_1\dots i_n}(t,x) \,,
\label{fdisc}
\eeq where the choices $\alpha=2,3$ will be most relevant, so that 
$\alpha$ is restricted to integer numbers hereafter.  
For convenience,  the main properties of  Laguerre polynomials are
reported in appendix \ref{app:A}. 
Using orthogonality, the coefficients $a^{(nk)}$ up to second order
read as follows:
 \bqa
\label{as}
\frac{m!}{(m+\alpha)!}\int d\bar{p}\,\bar{p}^\alpha \int \frac{d\Omega}{4\pi} f P^{(0)} L_m^{(\alpha)}
(\bar{p})&=& a^{(0m)}(t,x) \,,\nonumber\\
\frac{3 m!}{(m+\alpha)!}\int d\bar{p}\, \bar{p}^\alpha \int \frac{d\Omega}{4\pi} f P^{(1)}_i L_m^{(\alpha)}
(\bar{p})&=& a^{(1m)}_i(t,x) \,,\nonumber\\
\frac{15 m!}{2(m+\alpha)!}\int d\bar{p}\,\bar{p}^\alpha \int \frac{d\Omega}{4\pi} f P^{(2)}_{ij} L_m^{(\alpha)}
(\bar{p}) &=& a^{(2m)}_{ij}(t,x) \,,
\eqa where we have used the fact that $a^{(2m)}_{ij}$ may be taken to
be traceless and symmetric.

The ideal choice for the discretization of velocity space is once
again given by the requirement that the velocity moments
(\ref{tmunucons}) be represented \emph{exactly}. 
Use of non Cartesian coordinates,  however, implies 
that the associated lattice structure is no longer 
space-filling in general.

For a polynomial $P(\bar{p})$ of degree less than $2 N_p$, the
integral over $\bar{p}$ can be recast into an exact sum
$$
\int d\bar{p}\, {\bar p}\, e^{-\bar{p}} P(\bar{p}) =
\sum_{i=0}^{N_p-1} \omega^p_i\, P(\bar{p}_i) \,,
$$
if the nodes $\bar{p}_0,\ldots \bar{p}_{N_p-1}$ are given by
$L_{N_p}^{(\alpha)}(\bar{p}_i)=0$ and the weights $\omega_i$ are given
by
$$
\omega_i^p=\frac{(N_p+\alpha)!}{N_p!}\frac{\bar{p}_i}{(N_p+1)^2
  \left[L_{N_p+1}^{(\alpha)}(\bar{p}_i)\right]^2}\,.
$$
Note that the requirement (\ref{tmunucons}) implies a polynomial
$P(\bar{p})$ of degree $3-\alpha+N_p$, implying that $N_p>3-\alpha$ is a
necessary condition to represent this polynomial exactly.  
The choice of $\alpha$ may depend on the problem, but $\alpha=3$ is particularly
convenient here because it minimizes $N_p$.  
Unless one is interested in considering finite chemical potential, it is 
therefore convenient to choose $\alpha=3$, i.e. $N_p=1$, as we shall do from now on.

Coming to the angular dependence,  for polynomials $P$ of $\sin\phi$ and $\cos\phi$ 
of degree less than $2 N_\phi$, integration over the polar angle $\phi$ can be expressed as
$$
\int_{-\pi}^\pi d\phi\, P(\phi) = \sum_{l=0}^{2
  N_\phi-1}\omega_l^{\phi}\, P\left( \phi_l\right) \,,
$$
where $\phi_l=\frac{l \pi}{N_\phi}$ and
$\omega^\phi_l=\frac{\pi}{N_\phi}$.  
A polynomial $P^{(n)}(v^i)$ will only involve powers of $\cos\phi, \sin\phi$ up to $n$, so 
an exact representation of Eq.~(\ref{tmunucons}) requires $N_v\geq 3$, hence
$N_\phi\geq 3$.

Finally, for the integration w.r.t to the cosine of the polar angle
$\cos(\theta_p)=\xi$, there is a special symmetry that one can
exploit.  Namely, should the integrand contain a $p^x/p^0=\cos \phi
\sqrt{1-\xi^2}$, then the $\phi$-integration will only give a non-zero
value if another $\cos\phi$ is present in the integrand.  The only way
this can happen is through another factor $p^x/p^0$, which means
that $\cos^2\phi \left(1-\xi^2\right)$ must be present in the integrand. A
similar argument may be given for $p^y$. Hence, the integral over the
polar angle is always of the form $\int_{-1}^{+1} d\xi P(\xi)$, with
$P(\xi)$ a polynomial of degree $2+N_v$ or less. 
This is represented accurately as 
\beq
\label{disctheta}
\int_{-1}^{1}d\xi P(\xi) = \sum_{j=0}^{N_\xi-1} w^\xi_j P(\xi_j) \,,\eeq
for polynomial degrees less than $2 N_\xi$, $\xi_j$ being the
roots of the Legendre Polynomial $P_{N_\xi}(x)$ and 
\beq
\label{Legweight}
w_j^\xi=\frac{2}{(1-\xi_j^2)
\left(P_{N_\xi}^\prime(\xi_j)\right)^2}\,.  
\eeq
the corresponding weights.
   
The requirement (\ref{tmunucons}) again implies $N_\xi\ge 3$.



\begin{figure}[t]
\center
\includegraphics[width=0.8\linewidth]{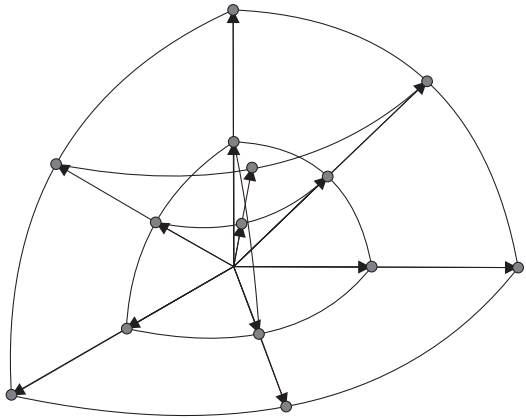}
\hfill
\caption{Example of a lattice configuration for the 
present relativistic model. 
The arrows denote the discrete spatial components of $p^\mu$.}
\label{fig:drel}
\end{figure}

To summarize, the requirements (\ref{tmunucons}) are fulfilled for the
ansatz (\ref{fdisc}) if one uses a discretized set of momenta
$$
{p_{ijl}}^{\mu} =T_0\bar{p_i}\left(
    \begin{array}{c}
      1\\
      \cos{\phi_l}\sqrt{1-\xi_j^2}\\
      \sin{\phi_l}\sqrt{1-\xi_j^2}\\
      \xi_j\\
    \end{array}
  \right) \,,
$$
with $\bar{p}_i$ the roots of $L_{N_p+1}^{(1)}$, $x_j$ the roots of
$P_{N_\xi}$ and $\phi_l=\frac{l \pi}{N_\phi}$, with $N_p,N_\xi,N_\phi$
greater or equal to $3$ and $3\leq N_v\leq{\rm min}(N_\phi,N_\xi)$
(see Fig.~\ref{fig:drel} for illustration). 

This implies a minimum number of $27$ discrete ``speeds'' ${p^\mu}_{ijl}$, quite
comparable with the number of discrete speeds commonly used
in non-relativistic LB theories.

With the momentum space thus discretized, we can move ahead and set up a concrete, fully
relativistic lattice Boltzmann algorithm.  

Before doing so, however,
it is instructive to study how to extract the fluid velocity and local
temperature distribution from a given distribution function $f\equiv f_{ijl}(t,x)$
discretized via (\ref{fdisc}).

\subsection{Energy-Momentum Conservation}

Using Eq.~(\ref{fdisc}), the energy momentum tensor becomes \beq
\label{tmunu}
T^{\mu\nu}=\frac{3T_{0}^4}{\pi^2}\left(
  \begin{array}{cc}
    a^{(00)}
    & a_i^{(10)}\\
    a_j^{(10)} &
    a_{ij}^{(20)}
    +\frac{1}{3}\delta^{ij}a^{(00)}
  \end{array}
\right) \,, \eeq

The requirement (\ref{tmunucons}) implies that $u^\mu$ should be a
future-pointing eigenvector of the energy momentum tensor, or $u^\mu
T_\mu^\nu=\epsilon u^\nu$. Thus, one has to calculate the eigensystem
of $T_\mu^\nu$ and identify $u^\mu$ as the (only) future-pointing
eigenvector, with eigenvalue $\epsilon$. Using existing numerical
packages, this can be done in a rather efficient way.  Once $\epsilon$
is known, the temperature is calculated from $\epsilon=\frac{3
  T^4}{\pi^2}$.  Since we neglected particle masses, the equation of
state is always that of an ideal gas, or $P=\frac{\epsilon}{3}$.
Non-ideal equations of state will be considered in a follow-up work.

\subsection{The equilibrium distribution function}

While the full equilibrium distribution function is given by
Eq.(\ref{Juttner}), the LB algorithm approximates all $f$'s by the
ansatz (\ref{fdisc}). 
Hence, for consistency, $f^{\rm eq}$ is also expanded as follows:
\bqa
\label{feq}
f^{\rm eq}(t,x,p) &=&e^{-\bar{p}}\left(a^{(00)}_{\rm eq}(t,x)+
P_i a^{(10)}_{i, \rm
eq}(t,x) +P_{ij} a^{(20)}_{ij, \rm eq}(t,x)\right.\nonumber\\
&&\left. +P_{ijk} a^{(30)}_{ijk, \rm
eq}(t,x)\right)\,, 
\eqa 
where we neglect higher order terms. 
The coefficients are readily evaluated to be \bqa
a^{(00)}_{\rm eq}&=& \left(1+\frac{4}{3}{\bf u}^2\right)\theta^{-4} \,, \nonumber\\
a^{(10)}_{i, \rm eq}&=&4 u^i u^0 \theta^{-4} \,, \nonumber\\
a^{(20)}_{ij, \rm eq}&=&
10 \left(u^i u^j-{\bf u}^2\frac{\delta_{ij}}{3}\right) \theta^{-4}\,, \nonumber\\
a^{(30)}_{ijk, \rm eq}&=&\frac{35}{12 {\bf u}^6}
P^{(3)}_{ijk}(u)  \bigg( u^0\left(15-10 {\bf u}^2+8 {\bf
u}^4\right) \nonumber \\ &-&\frac{15}{|{\bf u}|}\log\left( 1+2 {\bf u}^2+2 |{\bf
 u}|u^0\right) \bigg)\theta^{-4}\nonumber \,. \eqa 
      
With these expressions at hand, we are now ready to construct an operational algorithm.

\section{The fully relativistic LB algorithm}

First of all, a lattice version of the Boltzmann equation (\ref{BE})
needs to be established. 
For this purpose, we define three steps: 
i) Streaming in configuration space ($x$-move), ii) Streaming in momentum space ($p$-move),
iii) Collisional relaxation.
These read as follows (discrete indices are relaxed for notational simplicity):
\bqa
\label{lBE}
f^{\prime}(t,x^i,p^\alpha) &=& f\left(t+\delta t,x^i+\frac{p^i}{p^0}
  \delta t,p^\alpha\right) \,, \\
f^{\prime\prime}(t,x^i,p^\alpha)&=&
f(t,x^i,p^\alpha)+ \delta t\,\Gamma^\lambda _{\mu\nu} \frac{p^\mu p^\nu}{p^0}  \partial_\lambda^{(p)}f(t,x^i,p^\alpha)\,, \nonumber \\
f^\prime(t,x^i,p^\alpha)&=&f^{\prime\prime}(t,x^i,p^\alpha)-
\frac{p^\mu u_\mu \delta t}{p^0\tau_R}\left(f^{\prime\prime}-f^{\rm
    eq}\right)\,.\nonumber \eqa Here, we use the notation
$f(t,x^i,p^\alpha)  \equiv f(t,x,p)$ to make the vector components explicit.

The first step consists in the streaming of the distribution functions
according to the discrete momenta. The second one, is the
implementation of the external forces due to the curvature of the
space-time, and the third one is the collision step, expressed
in terms of relaxation towards the local equilibrium. 

The relativistic lattice Boltzmann algorithm is given by the
following sequence of steps:
\begin{enumerate}

\item {\it Initialization}: 

At the initial step, $f^\prime$ needs to be known at grid sites
$x^i$ and discretized momenta $p^\alpha$. 
For this, we introduce the initial conditions of the specific problem
and the initial distributions is typically specified as the local
equilibria associated with the initial hydrodynamic fields
$f(t=t_0,x^i,p^{\alpha}) = f^{eq}(t,x,p)$, where the equilibrium distribution
is given by the expression (\ref{feq}).
 
\item {\it $x$-move}: 

Calculate the new $f$ from $$
f(t,x^i,p^\alpha)=f^\prime\left(t-\delta t, x^i-v^i \delta t, p^\alpha
\right).
$$In non-relativistic lattice Boltzmann methods, this step is known as streaming, each
populations moves to the site pointed by its corresponding discrete speed.

\item {\it $p$-move}: 

Compute the change in the distribution function because of external/internal forces.
In principle, a force term implies a change in velocity, thereby jeopardizing
the discrete structure of velocity space.
However, this can be preserved by moving the information according to
the streaming step given above (with constant speeds/momenta) and then
{\it correcting} the distribution form with an appropriate source term. 
The latter is identified by representing the derivative in momentum space
also as a polynomial expansion:
 \beq
 S(t,x) \equiv  F^\lambda \partial_\lambda^{(p)}f =e^{-\bar{p}} \sum_{n=0}^{N_v-1}
  P^{(n)}_{i_1\ldots i_n}({\bf v}) s^{(n)}_{i_1\dots i_n}(t,x)
\label{explicitpmove}
\,, \eeq 
with $F^\lambda = \Gamma^\lambda _{\mu\nu} \frac{p^\mu
 p^\nu}{p^0}$.
 
The unknown source coefficients $s^{(n)}$ can be computed by inverting
(\ref{explicitpmove}) and using integration by parts.
Note that in general these coefficients can be
expressed as sums over the coefficients $a^{(n0)}$ in Eq.~(\ref{fdisc}).  

With the source term available, we evaluate $f^{\prime\prime}$ as 
\bqa
f^{\prime\prime}(t,x^i,p^\alpha) &=& f(t,x^i,p^\alpha) \nonumber \\ &+& \delta t\,
e^{-\bar{p}} \sum_{n=0}^{N_v-1} P^{(n)}_{i_1\ldots i_n}({\bf v})
s^{(n)}_{i_1\dots i_n}(t,x) \nonumber \,. 
\eqa 
This step accounts for the geometric forces in the given space-time.

\item {\it Equilibration}: 

In order to perform collisional relaxation, local discrete equilibria 
must be constructed first.
To this purpose, we calculate the energy momentum tensor
 $T^{\mu\nu}$ corresponding to $f^{\prime\prime}$ from
 Eqs.~(\ref{tmunu}) and (\ref{fdisc}). We then compute the values of the fluid
 $4$-velocity $u^\mu$ and energy density $\epsilon$ by calculating the
 future-pointing eigenvector of $T^{\mu\nu}$. 
 The local temperature is obtained by the equation of state.  
 Using $T,u^\mu$, we calculate $f^{\rm eq}$ from Eq.~(\ref{feq}), and 
 the collision coefficient $\Omega=p^\mu u_\mu/(p^0 \tau_R)$.

\item {\it Collision}: 

Calculate the post-collisional state $f^\prime$ from the known
values of $f^{\prime\prime}$, $f^{\rm eq}$, and $\Omega$, according
to 
\bqa
 f^\prime(t,x^i,p^\alpha)&=&f^{\prime\prime}(t,x^i,p^\alpha)(1-
 \Omega \delta t) \nonumber \\ &+&\Omega \delta t f^{\rm eq}
 (t,x^i,p^\alpha) \,.
\label{collide}
\eqa
\item Cycle through $2-5$ for each time-step until completion of 
the time evolution.
\end{enumerate}

A remark concerning the $x$-move step is in order. 
Since the unit vectors $v^i=p^i/|{\bf p}|$ discretize the unit sphere (see
Fig.~\ref{fig:drel}), the displaced positions
$x^i-v^i \delta t$ will typically not correspond
to a neighbouring grid site, unless very particular
geometries (e.g. a hexagonal lattice) are chosen. 
This breaks the light-cone rule discussed previously.

At least two ways out of this problem can be envisaged. 

The first is to acknowledge the fact that spatial and momentum discretization
can no longer be kept in synchrony, going back to the original Boltzmann
equation (\ref{BE}) and discretize space derivatives in a
flux-conserving way, according to one's favored finite-volume/difference practice. 
This is similar to the non-relativistic LB method
on so-called unstructured meshes, wherein powerful features of modern
finite-volume techniques are imported within the LB framework. 
For more details about this technique, we refer the reader to the
literature, e.g. Refs.~\cite{ulbm1, ulbm2}.
The second method, the one we will adopt in the following, consists of transferring
off-grid distributions into grid locations through (bilinear) interpolation.

Either way, it is clear that none of the two methods above can match
the simplicity, hence computational efficiency, of the space-filling
Cartesian formulation. In particular, they cannot preserve the exact
nature of the streaming step in light-cone form.
This limitation appears to be inherent to the
non-separability of the relativistic J\"uttner equilibria 
along the three spatial coordinates. 
In this respect, a complete transfer of the key assets of the 
non-relativistic scheme, does not appear to be feasible.
This is the price to pay for a fully relativistic lattice formulation.
However, since information is still traveling along constant streamlines
in configuration space, this does not prevent the fully relativistic 
LB algorithm from delivering competitive performance, as we shall demonstrate in the
next section, where concrete test-case simulation are presented.

\section{Application: The boost-invariant Quark-Gluon Plasma}

As a first application, let us consider the Milne spacetime generated by
the coordinate transformation $\tau=\sqrt{t^2-z^2}$, $Y={\rm
tanh^{-1}}(z/t)$.  In these coordinates, the metric takes the form
$$
g_{\mu \nu}={\rm diag}(g_{\tau \tau},g_{xx},g_{yy},g_{YY})={\rm
diag}(1,-1,-1,-\tau^2)\,,
$$
where hereafter the $(+,-,-,-)$ sign convention is assumed.
In this metric, the non-vanishing Christoffel symbols are:
$$
\Gamma_{\tau T}^{Y}=\Gamma_{Y
  \tau}^{Y}=\frac{1}{\tau}\,,\quad \Gamma^\tau_{YY}=\tau\,.
$$
This implies a non-vanishing covariant fluid divergence even for a fluid a rest,
$u^\mu=(1,\vec{0})$,  i.e.
$$
\nabla_\mu u^\mu=\partial_\mu u^\mu + \Gamma^\mu_{\mu \nu} u^\nu =
\Gamma^Y_{Y \tau} u^\tau=\frac{1}{\tau}\neq 0\,.
$$

The reason for this is that the Milne space-time is expanding in one
dimension, so that a system at rest experiences the 'stretching' of
space-time. This is a nice feature, because it naturally mimics the expansion
of the system, following a heavy-ion collision in Minkowski space (see
Ref.~\cite{Romatschke:2009im}, Section 5B for details).  
In Minkowski space-time, a general solution to $p^\mu \partial_\mu f=0$ is e.g.
$f=f(p_\perp,t \vec{p}- \vec{x} p^t)$. 
In Milne-coordinates that would
correspond to 
\beq f=f(p_\perp,\tau^2 p^Y,\cosh{Y} \left(\tau
 \vec{p}_\perp-\vec{x}_\perp p^\tau\right)-\tau p^Y \vec{x}_\perp
\sinh{Y})\,.
\label{fullfstream}
\eeq 
This can be further simplified by considering only the evolution
at mid-rapidity, $Y\simeq 0$. It is readily checked that, under such 
condition, the action of the derivative $\partial_\tau^{(p)}$ is exactly cancelled by the
$\partial_Y$ derivative of the last term in (\ref{fullfstream}).
One may thus neglect both, so that the Boltzmann equation at mid-rapidity
simplifies to
\beq
\label{BEreduced}
\left[p^\tau \partial_\tau+\vec{p}_\perp \cdot
  \vec{\partial}_\perp\right]f -2 \frac{\partial f}{\partial p^Y}
\Gamma^Y_{Y \tau} p^{Y} p^\tau = - {\cal C}[f]\,,\qquad
Y\simeq 0\,, 
\eeq where $p^\tau$ is treated as an independent
variable (e.g. $\partial_Y^{(p)}p^\tau=0$).  
The corresponding discrete version of the lattice Boltzmann 
equation takes the following form:
\bqa
\label{BEMilnedisc}
f\left(\tau+\delta \tau,x^i+\frac{p^i}{p^\tau} \delta \tau, p^\alpha
\right)
&=& f^\prime(\tau,x^i,p^\alpha)\,,\\
f^{\prime\prime}(\tau,x^i,p^\alpha)&=&f(\tau,x^i,p^\alpha)+\frac{2\delta \tau}{\tau} p^Y \partial_Y^{(p)}f \,, \nonumber\\
f^\prime(\tau,x^i,p^\alpha)&=&f^{\prime\prime}(\tau,x^i,p^\alpha)
\nonumber \\ &-& \frac{p^\mu u_\mu \delta
  \tau}{p^\tau\tau_R}\left(f^{\prime\prime}-f^{\rm eq}\right)
\nonumber \,. \eqa

\subsection{Warmup: 0+1dimensions in Milne space-time}

The simplest practical example is given by considering a system that
is homogeneous in the Milne coordinates, e.g.  $f = f(\tau,p^\alpha) =
f(\tau,\bar{p},\xi_i)$.  Because there is no space-dependence left,
one may use a simplified version of the discretization ansatz
(\ref{fdisc}), \beq f(\tau,p^\alpha)=e^{-4} \sum_{k=0}^{N_\xi-1}
a_{0k}(\tau) P_k\left(\xi\right)\,, \eeq 
where $p^z=p^Y \tau$, such
that $p^\tau=\sqrt{p_\perp^2+\left(p^{Y}\tau\right)^2}$ becomes
$p^\tau=|{\bf p}|=\sqrt{p_\perp^2+p_z^2}$ and $\xi={\rm
arccos}\frac{p^z}{p^\tau}$.  
With $f$ discretized this way, one readily inverts to obtain the 
coefficients $a_{ml}$. 
For instance
\bqa
\label{amlMilne}
a_{0l}&=&\frac{2l+1}{12}\int_0^\infty d\bar{p}\, \bar{p}^3 \int d\xi
P_l(\xi) f\left(\tau, \bar{p}, \xi\right)\nonumber\\
&=&\frac{2l+1}{2} e^{4}\sum_{j=0}^{N_\xi-1}\omega_j^\xi P_l(\xi_j)
f(\bar{p}_i,\xi_j) \,, \eqa where the second line is the discretized
version and we recall that $\xi_j$ are the roots of $P_{N_\xi}$ and
the weights $\omega^\xi$ were specified above.
Starting with an initial equilibrium distribution function
$$
f(\tau_0,\bar{p},\xi_i) \propto e^{-\bar{p}/\theta}\,,
$$
and discretizing $f(\tau_0)$ on a $\xi$ grid with $N_\xi$ points
($\bar p=4$), the algorithm reads as follows.

First set $f^\prime = f$. 
Then, set
$$
p^z\partial_{z}^{(p)}f=e^{-\bar p}\left(S^0+S^2 P_2(\xi)\right) \,,
$$
where the coefficients are calculated to be\bqa
S^0&=&-\frac{1}{12}\int d\bar{p} {\bar p}^3 \int d\xi \left(1+\xi^2\right) f \,, \nonumber\\
S^2&=&-\frac{5}{24}\int d\bar{p} {\bar p}^3 \int d\xi
\left(3\xi^4-8\xi^2+1\right) f\,.  \nonumber \eqa Calculate
$f^{\prime\prime}$ from
$$
f^{\prime\prime}(\tau_0+\delta\tau, \bar{p}, \xi_i)=f(\tau_0,\bar
p,\xi_i)+\frac{\delta \tau}{\tau+\delta \tau} e^{-\bar p}\left(S^0+S^2
  P_2(\xi)\right) \,.
$$
Next, calculate the equilibrium temperature via the energy-momentum
tensor corresponding to $f^{\prime\prime}$ and obtain
$f^\prime(\tau_0+\delta \tau)$ from Eq.(\ref{collide}). 

The above  steps are cycled in time over the prescribed time-span
of the simulation.   

\subsection{Comparison with exact results}
\label{sec:52}

Let us now compare the results of the above lattice Boltzmann
algorithm in Milne-spacetime against known exact results. 
First, let us consider the case in which the 
relaxation time is so large that the collision term plays a negligible role.
The solution to the Boltzmann equation must therefore be very
close to the free streaming solution,
$$
f_{\rm free-stream}(\tau,p,\xi)=e^{-p \sqrt{1+\xi^2 r}/Q}\,,
$$
where  $r=\frac{\tau^2}{\tau_0^2}-1$ and $Q$ is the initial
temperature scale. From this, the temperature can be obtained from
the energy-momentum tensor as
$$
T(\tau)=Q\left[\frac{1}{2}\left(\frac{\tau_0^2}{\tau^2}+\frac{
      \arctan{\sqrt{\tau^2/\tau_0^2-1}}}{\sqrt{\tau^2/\tau_0^2-1}}\right)\right]^{1/4}
\,,
$$
As opposed to the free-streaming case, let us now consider the 
opposite extreme of small viscosity, i.e. very fast relaxation to the
local equilibrium. 
In this case, the 'exact' temperature evolution is given by fluid dynamics. 
More specifically, denoting $T^Y_Y-P\equiv\Phi$, the energy density and $\Phi$
fulfill the coupled equations \cite{Romatschke:2009im} \bqa
\label{hydroeq}
\partial_\tau \epsilon &=& -\frac{\epsilon+P}{\tau}+\frac{\Phi}{\tau}\,,\nonumber\\
\partial_\tau \Phi &=& -\frac{\Phi}{\tau_\pi}+\frac{4 \eta}{3 \tau_\pi
  \tau}-\frac{4\Phi}{3 \tau}-\frac{\lambda_1}{2 \tau_\pi
  \eta^2}\Phi^2\,, \eqa where $\tau_\pi$ is the relaxation time and
$\lambda_1$ is a self-coupling parameter that can be calculated
\cite{York:2008rr}, presumably fairly easily for the BGK collision
kernel used here. 

Since both are second-order corrections to
hydrodynamics, their determination is left for future work. 
Here, we simply set $\lambda_1=\frac{6 \eta^2}{(\epsilon+P)}$ and vary
$\tau_\pi$ between $\tau_\pi=\frac{2(2-\ln 2)}{T}\frac{\eta}{s}$ and
$\tau_\pi=\frac{6}{T}\frac{\eta}{s}$ (the weak and strong coupling
limit, respectively, \cite{Baier:2007ix}).  The above set of hydrodynamic
equations display simple analytic solutions for the case of vanishing
viscosity (ideal fluid) and first order gradient expansion
(Navier-Stokes). These are \cite{Muronga:2001zk,Baier:2006um} given by
\beq T(\tau)=T_0\left(\frac{\tau_0}{\tau}\right)^{1/3}\left[
  1+\frac{2\eta}{3 s \tau_0
    T_0}\left(1-\left(\frac{\tau_0}{\tau}\right)^{2/3}
  \right)\right]\,.  \eeq

The full set of equations (\ref{hydroeq}) is second-order in gradients
and thus causal for $\tau_\pi$ larger than some critical value.  
For general values of $\tau_\pi,\lambda_1$, it has to be solved
numerically.

\begin{figure}[t]
\center
\includegraphics[width=0.9\linewidth]{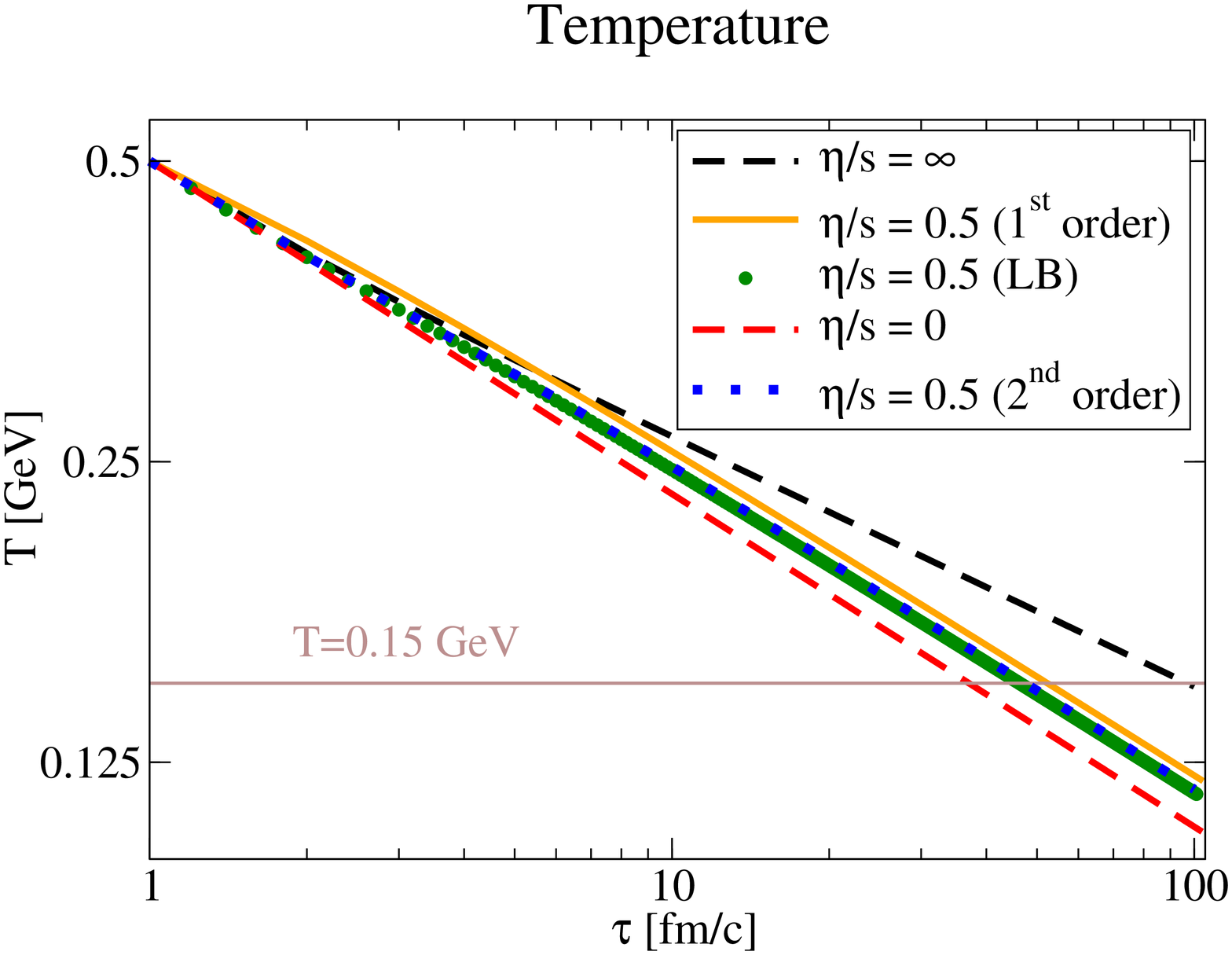}
\hfill
\includegraphics[width=0.9\linewidth]{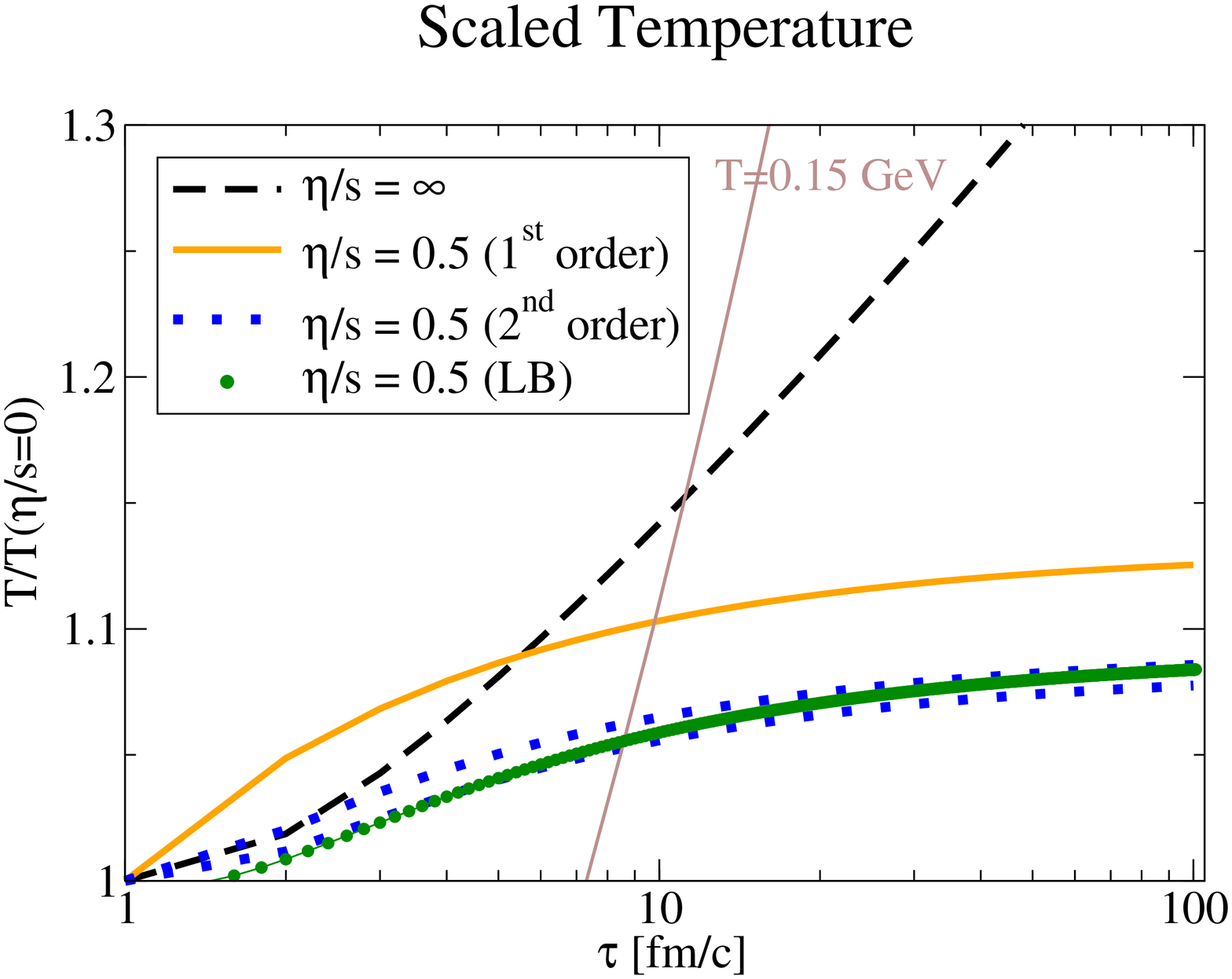}
\caption{Temperature evolution of a system starting at $\tau_0=1$ fm/c
and temperature $T_0=0.5$ GeV. Results shown are $1^{\rm st}$ order
viscous fluid dynamics (Navier-Stokes) (initially not in
equilibrium), $2^{\rm nd}$ order viscous fluid dynamics 
(two values of $\tau_\pi$, see text), and the lattice Boltzmann
result (LB) for $N_\xi=5$, all for $\eta/s=0.5$.  
As a reference,  the free streaming result 
($\eta/s \rightarrow \infty$) and the result for ideal fluid
dynamics ($\eta/s=0$) are also shown (top panel).  
Bottom panel: results are divided by the ideal fluid dynamic result 
to highlight differences.
The LB result is seen to converge to the $2^{\rm nd}$ order 
viscous fluid dynamics.
}
\label{fig:comp1}
\end{figure}


In Fig.\ref{fig:comp1}, we show a comparison between the lattice
Boltzmann algorithm, against various 'exact' results for the case
of $\eta/s=0.5$ and $T_0(\tau_0=1{\rm fm/c})=0.5$ GeV.
  

The second-order set of hydrodynamic equations was solved using forward time
differencing $\delta \tau \partial_\tau X(\tau)=\left(X(\tau+\delta
  \tau)-X(\tau)\right)$.  A time step of $\delta \tau=0.01 \tau_0$
was required to reach a stable continuum result.  Conversely,
for the lattice Boltzmann algorithm, typically the result is stable
for $\delta \tau<0.2 \tau_0$, nearly $20$ times larger than the fluid
dynamics requirement and only about $5$ times smaller than standard LB schems
in cartesian geometries. 
Based on the general arguments discussed earlier on in this paper, we
interpret this encouraging outcome as the beneficial effect
of moving information along constant streamlines. 

As can be seen from this figure, the lattice
Boltzmann algorithm does track the $2^{\rm nd}$ order viscous fluid
dynamics result from early times onwards. (The Navier-Stokes result
has a different initial condition and hence the other results are not
expected to track this curve).  The numbers in Fig. \ref{fig:comp1}
were chosen such that the initial temperature corresponds to the
maximum temperature expected for heavy-ion collisions at
$\sqrt{s}=5.5$ TeV at the Large Hadron Collider. 
Moreover, $T=0.15$ GeV is the temperature where a freeze-out 
to hadrons is expected. Hence, for this temperature region, the lattice Boltzmann 
algorithm with $N_\xi=5$ accurately reproduces the $2^{\rm nd}$ order 
viscous fluid dynamics results.
\begin{figure}[t]
\center
\includegraphics[width=0.9\linewidth]{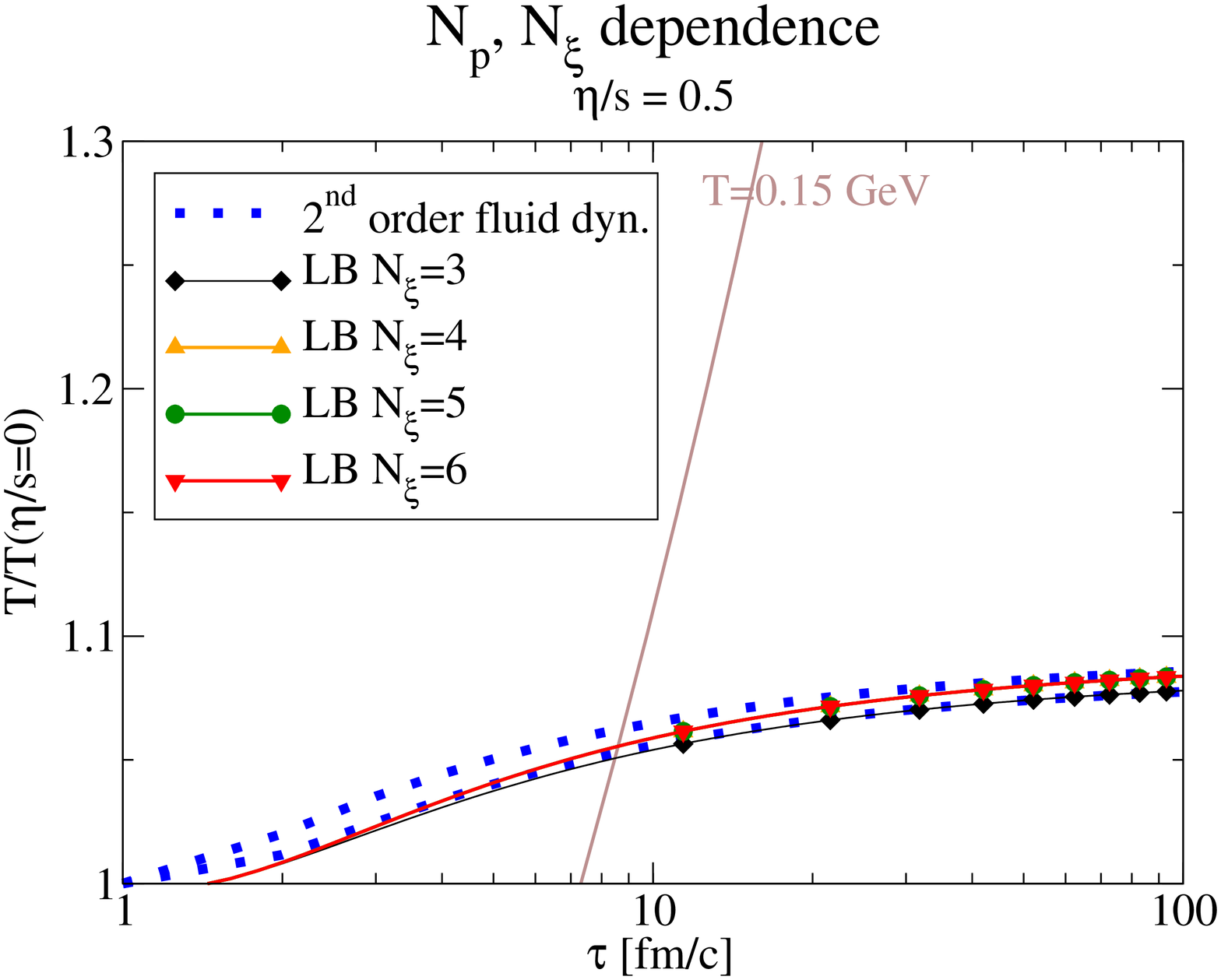}
\hfill
\includegraphics[width=0.9\linewidth]{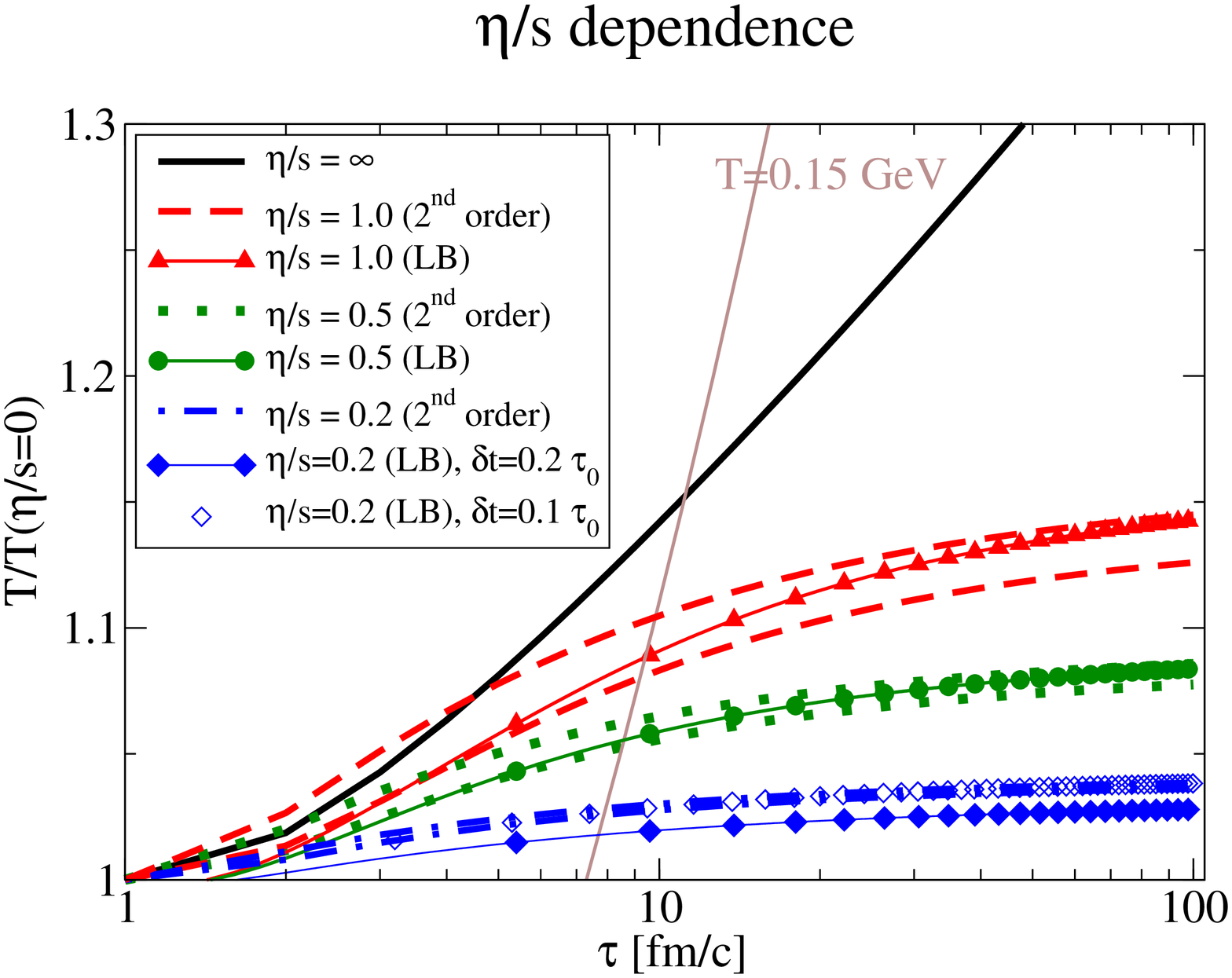}
\caption{Top: dependence of temperature evolution on discretization
(for $\eta/s=0.5$).
Bottom: dependence of result on value of shear viscosity coefficient
compared to $2^{\rm nd}$ order viscous fluid dynamics (for $N_\xi=5$).
Viscous fluid results from (\ref{hydroeq}) for two values of $\tau_\pi$, see text.
For the lowest viscosity shown ($\eta/s=0.2$) we highlight the effect
of numerical viscosity by choosing two different values of $\delta \tau$.}
\label{fig:comp2}
\end{figure}

The dependence of the lattice Boltzmann result on the chosen
discretization is shown in Fig.\ref{fig:comp2}. 
There, one can see that all cases (even $N_\xi=3$) reproduce the viscous fluid result,
and for finer discretization $N_\xi\ge 4$, the LB results are
indistinguishable to the naked eye.
The viscosity dependence is also shown in Fig.\ref{fig:comp2}.  
The lattice Boltzmann algorithm tracks the fluid dynamic result for
viscosities up to $\eta/s\sim 1.0$. 
For smaller viscosities, $\eta/s < 0.5$, 
LB  undershoots the fluid dynamics result. 
However, decreasing $\delta \tau$ by a factor two, brings the 
LB back in line with the fluid dynamic result.

\subsection{2+1 dimension in Milne space-time}

Next we consider the case in which the transverse dynamics
is also taken account. 
Inclusion of the transverse space dynamics requires the solution of
Eq.~(\ref{BEMilnedisc}), using the full discretization (\ref{fdisc}).
For convenience, we choose a square lattice for the grid in $x_\perp$.
Choosing furthermore $\delta \tau=\delta x$, we use bilinear
interpolation to obtain $f^\prime$ at points that lie in-between lattice sites.
The change in momenta $p$ is calculated similar to Eq.(\ref{explicitpmove}).

The LB solver is applied to simulate the evolution of the medium
created in $Au+Au$ collisions at top RHIC energies ($\sqrt{s}=200$
GeV).  For smooth initial conditions, the results may then be compared
to the fluid dynamics solution, given by the code VH2+1
\cite{Romatschke:2007mq,Luzum:2008cw}.  This code has been
cross-tested against several other codes and is generally credited 
for producing reliable results for smooth initial conditions.

The initial conditions are generated at initial time $\tau=1$ fm/c from
a Glauber model, with number of collisions scaling
(c.f.\cite{Romatschke:2009im}) on a $69\times 69$, $139\times 139$ or
$279\times 279$ grid \ with a lattice spacing of $\delta x=0.4$ fm,
$\delta x=0.2$ fm or $\delta x=0.1$ fm, respectively.  The maximum
temperature at the center of the grid is $T^{\rm max}=0.37$ GeV for
central collisions (impact parameter $b=0$ fm).

\begin{figure}[t]
\label{fig:b7}
\center
\includegraphics[width=0.9\linewidth]{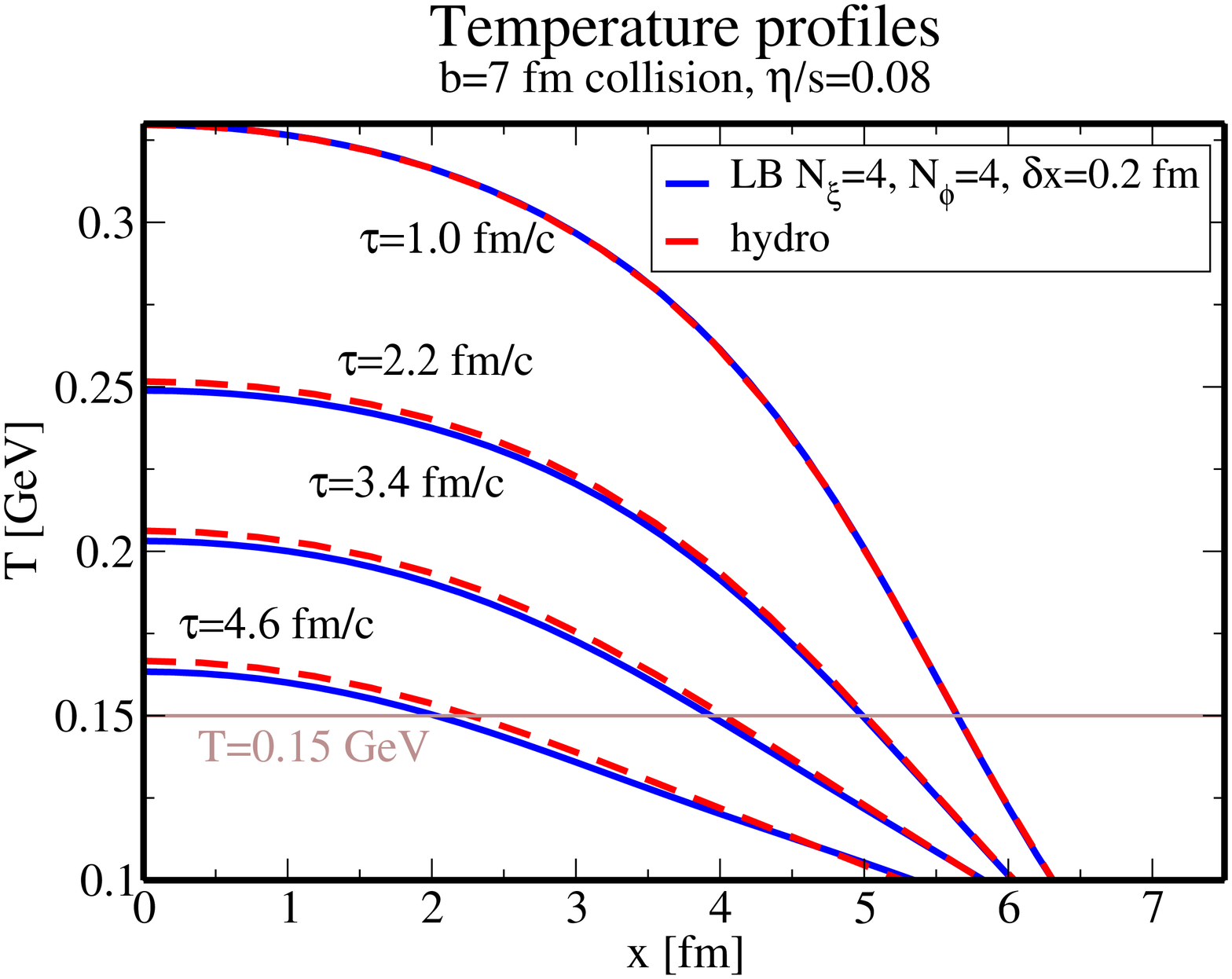}
\hfill
\includegraphics[width=0.9\linewidth]{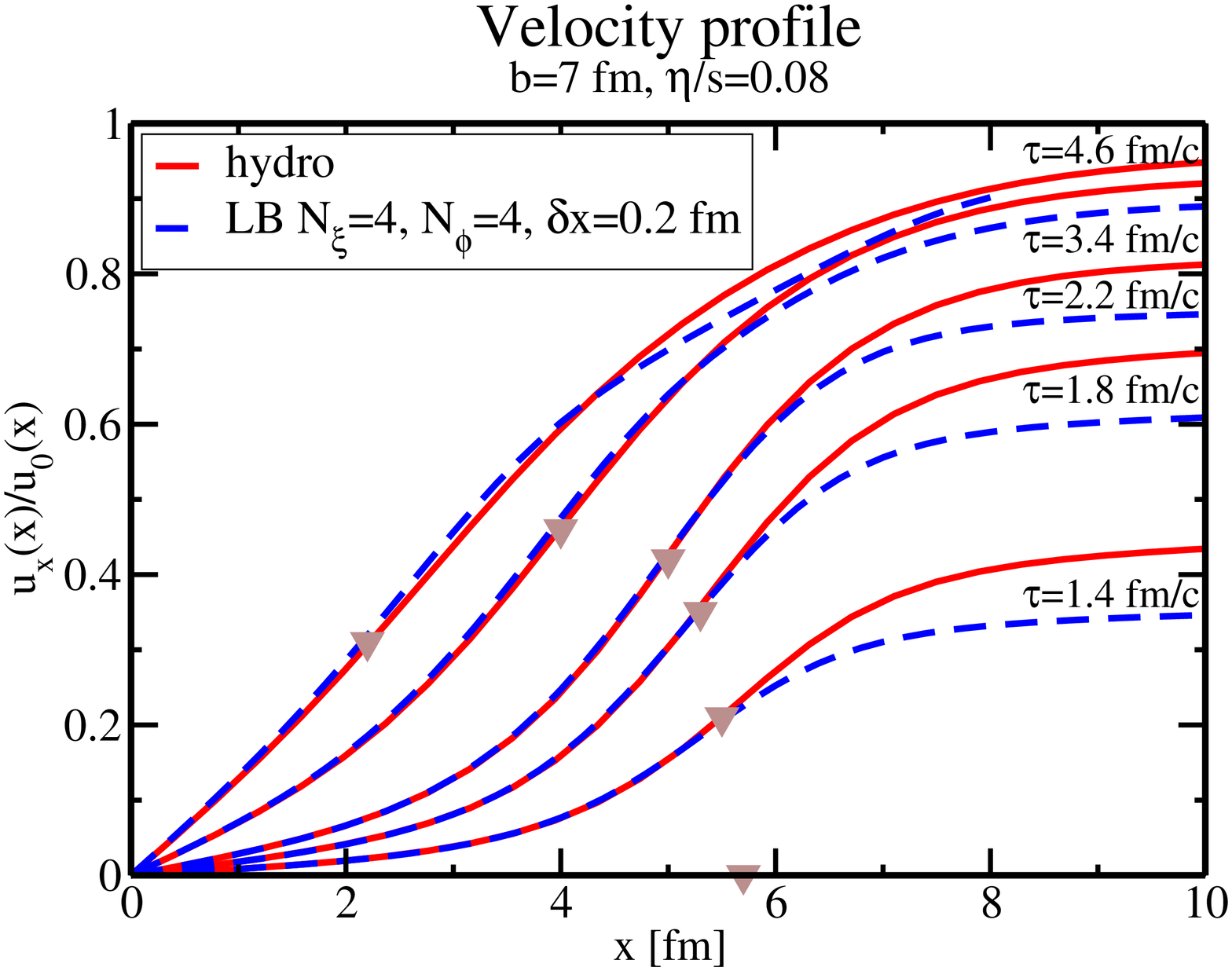}
\caption{Top: temperature evolution in viscous hydrodynamics
('hydro') versus lattice Boltzmann equation ('LB') for
$\eta/s=0.08$.  As can be seen, the temperature evolution in the
lattice Boltzmann approach for $\delta x=0.2$ fm is reasonably close
to the 'exact' hydrodynamic result.  Bottom: evolution of velocity $u^x/u^0$ in
viscous hydrodynamics versus LB. Even high velocities up to
$80$ percent of the speed of light are well represented, but at later times an instability
develops in the low temperature region (a grey triangle marks the
position of $T=0.15$ GeV).  Note that the discrepancy at larger $x$
lies exclusively in the region $T<0.15$ GeV (compare top plot).}
\end{figure}


In Fig.~\ref{fig:b7}, the temperature evolution from LB with
$N_\xi=4,N_\phi=4$ for $\delta x=0.2$ fm is compared to the
VH2+1 solution. 
As can be seen from this figure, the LB code reproduces the
VH2+1 result rather accurately for temperatures $T\ge 0.15$ GeV, the
relevant temperature regime for the fluid medium. 
At later times $\tau>5 fm/c$, when all fluid cells have cooled below a temperature of
$0.15$ GeV, numerical instabilities develop at the outer edges, for the
discretization used $N_\xi=4,N_\phi=4$. We have checked that the
remaining discrepancy between fluid dynamics and LB in the temperature
evolution close to $x\sim 0$ can be cured by using a smaller lattice
spacing $\delta x=0.1$ fm.

Also shown in Fig.~\ref{fig:b7} is the evolution of the velocity
$\frac{u_x}{u_0}$. One can see that LB tracks the VH2+1 result
closely for the high temperature region (there are clear deviations at
low temperatures $T<0.15$ GeV, compare left plot in
Fig.~\ref{fig:b7}). Even high velocities seem to be well represented,
but the algorithm does not handle correctly velocities $u_x/u_0>0.8$.
Presumably, this discrepancy can be cured by including higher order
terms in Eq.(\ref{feq}). However, it should be pointed out that for
the high temperature region $T>0.15$, the velocity distributions from
fluid dynamics are accurately reproduced.

\section{Conclusions}

Summarizing, we have developed a new scheme based on the
lattice-Boltzmann method to model relativistic fluid dynamics in
general spacetime. 
The main advantage of our scheme, as compared with previous
relativistic lattice Boltzmann models \cite{RELB1, RELB2}, rests mainly
with its ability describe the dynamics of ultra-relativistic systems in
general space-time geometries. 
The present model differs from typical lattice Boltzmann schemes mostly
in the streaming step, which, because of the spherical shape
of the discrete momenta, is no longer space-filling.
Instead, multi-linear interpolation is used to represent the distribution functions 
in the second-nearest neighbours of each cell on the lattice.
This interpolation breaks the {\it exact} nature of the standard LB streaming 
operator. However, at variance with hydrodynamic formulations, it
still moves information along constant streamlines, thereby permitting
to march in larger time-steps than hydrodynamic codes. 
Our scheme has been validated through simulations in quark-gluon
plasma, yielding very satisfactory agreement with other computational methods
based on a macroscopic description, at a lower computational cost
(nearly two orders of magnitude faster than the VH2+1 viscous hydro code 
\cite{Romatschke:2007mq} and still a factor $3-5$ as compared to optimized ones \cite{perscomm}).

Because of these favorable properties, we expect this new LB method
to offer a new competitive entry for the computational study 
of large-scale complex relativistic fluids.

\section*{Acknowledgements}

We would like to thank G.~Denicol and H.~Herrmann for interesting
discussions. This work was supported in part by the Helmholtz International
Center for FAIR within the framework of the LOEWE program launched
by the state of Hesse.

\begin{appendix}

\section{The $P^{(n)}$ Legendre polynomials}
\label{sec:B}

The vector polynomials $P^{(n)}_{i_1\ldots i_n}({\bf v})$ are
constructed by requiring orthogonality with respect to the angular
integral $\int \frac{d\Omega}{4\pi}$. One finds Specifically, the
polynomials involved are given by \bqa
P^{(0)}&=&1\,,\nonumber\\
P^{(1)}_i&=&v_i\,,\nonumber\\
P^{(2)}_{ij}&=&v_i v_j - \frac{1}{3}\delta_{ij} \,,\nonumber\\
P^{(3)}_{ijk}&=&v_i v_j v_k-\frac{1}{5}\left(\delta_{ij}
  v_k+\delta_{ik} v_j
  +\delta_{jk} v_i\right) \,,\nonumber\\
&\ldots& \eqa The orthogonality relations for the first few
polynomials are found to be \bqa
\int \frac{d\Omega}{4\pi} P^0 P^0&=&1\,,\nonumber\\
\int \frac{d\Omega}{4\pi} P^{(1)}_i P^{(1)}_j&=&\frac{\delta_{ij}}{3}\,,\nonumber\\
\int \frac{d\Omega}{4\pi} P^{(2)}_{ij} P^{(2)}_{lm}&=&\frac{1}{15}\left(\delta_{il}\delta_{jm}+\delta_{im}\delta_{jl}-\frac{2}{3}\delta_{ij}\delta_{lm}\right)\,,\nonumber\\
&\ldots& \eqa

\section{Generalized Laguerre Polynomials}
\label{app:A}

The first few generalized Laguerre Polynomials are given by \bqa
L_0^{(\alpha)}(x)&=&1 \,, \\
L_1^{(\alpha)}(x)&=&1+\alpha-x \,, \nonumber\\
L_2^{(\alpha)}(x)&=&\frac{x^2}{2}-(\alpha+2)x
+\frac{(\alpha+2)(\alpha+1)}{2} \,,
\nonumber\\
L_3^{(\alpha)}(x)&=&-\frac{x^3}{6}+\frac{(\alpha+3)
  x^2}{2}-\frac{(\alpha+3)(\alpha+2) x}{2} \nonumber
\\&+&\frac{(\alpha+3)(\alpha+2)(\alpha+1)}{6}\,.  \nonumber \eqa

The orthogonality relation is given by \beq \int_0^\infty dx\ e^{-x}\
x\ L_n^{(\alpha)}(x) L_m^{(\alpha)}(x)=
\frac{\Gamma(n+\alpha+1)}{n!}\delta_{nm} \,. \eeq

\end{appendix}

\end{document}